# Cancer classification and pathway discovery using non-negative matrix factorization

Zexian Zeng
Preventive Medicine
Northwestern University
Feinberg School of Medicine
Chicago, IL, USA
zexian.zeng@northwestern.edu

Andy Vo
Developmental Biology and
Regenerative Medicine
University of Chicago
Chicago, IL, USA
ahvo@uchicago.edu

Chengsheng Mao
Preventive Medicine
Northwestern University
Feinberg School of Medicine
Chicago, IL, USA
chengsheng.mao@northwestern.edu

Susan E Clare*
Surgery
Northwestern University
Feinberg School of Medicine
Chicago, IL, USA
susan.clare@northwestern.edu

Seema A Khan*
Surgery
Northwestern University
Feinberg School of Medicine
Chicago, IL, USA
s-khan2@northwestern.edu

Yuan Luo*
Preventive Medicine
Northwestern University
Feinberg School of Medicine
Chicago, IL, USA
yuan.luo@northwestern.edu

*Abstract*— **Extracting genetic information from a full range of sequencing data is important for understanding diseases. We propose a novel method to effectively explore the landscape of genetic mutations and aggregate them to predict cancer type. We used multinomial logistic regression, nonsmooth non-negative matrix factorization (nsNMF), and support vector machine (SVM) to utilize the full range of sequencing data, aiming at better aggregating genetic mutations and improving their power in predicting cancer types. Specifically, we introduced a classifier to distinguish cancer types using somatic mutations obtained from whole-exome sequencing data. Mutations were identified from multiple cancers and scored using SIFT, PP2, and CADD, and grouped at the individual gene level. The nsNMF was then applied to reduce dimensionality and to obtain coefficient and basis matrices. A feature matrix was derived from the obtained matrices to train a classifier for cancer type classification with the SVM model. We have demonstrated that the classifier was able to distinguish the cancer types with reasonable accuracy. In five-fold cross-validations using mutation counts as features, the average prediction accuracy was 77.1% (SEM=0.1%), significantly outperforming baselines and outperforming models using mutation scores as features. Using the factor matrices derived from the nsNMF, we identified multiple genes and pathways that are significantly associated with each cancer type. This study presents a generic and complete pipeline to study the associations between somatic mutations and cancers. The discovered genes and pathways associated with each cancer type can lead to biological insights. The proposed method can be adapted to other studies for disease classification and pathway discovery.**

## I. INTRODUCTION

Understanding the association between genetics and disease is important for understanding the underlying pathophysiology, planning treatment, and predicting prognostic outcomes. In cancer, many molecular and genomic studies have identified somatic mutations within numerous genes associated with cancer initiation, progression, and treatment responses [1-3]. More recently, the idea of personalized medicine is becoming increasingly popular where genetic profiles of tumors can be used to guide clinical decisions such as treatment options and preventive measures [4]. The development of massively parallel, high throughput DNA sequencing technology has enabled the cataloging of somatic mutations in cancer, making personalized medicine increasingly achievable.

The majority of genome-wide sequencing studies have focused on the identification of individual driver genes [5]. However, driver mutations are often highly heterogeneous between cancer genomes, even within the same type [6]. Furthermore, studies have observed cancer to be highly complex, often resulting from multiple interacting mutations and related pathways [7, 8]. While many methods attempt to address the complex mutational heterogeneity in disease, it still remains challenging due to limited study-power and lack of complete knowledge regarding gene and pathway interaction, particularly in cancer [9-13]. Individually, genes can show a small contribution to disease phenotype but can have a significant effect when analyzed together [14]. Thus, it is important to consider methods that can encompass the full scope of genes. When genes and mutations are studied together, novel biological interactions and pathways can be identified,

* indicates co-corresponding authors. This study was supported in part by grant R21LM012618-01 from the NIH.

and the information can further provide us with biological and clinical insights.

## II. RELATED WORK

Numerous studies have been performed for cancer classification using information from somatic mutations. Early work mainly relied on gene expression profiles for cancer type classification [15-17]. PAM50 is a well-known panel for cancer subtype classification, which utilizes 50 gene expression levels and classifies breast cancer mainly into four subtypes. Recently, DNA methylation profiles have been explored [18, 19] for cancer type classification. The majority of studies involving DNA mutations have used mutations as individual variables. Soh et al. [20] used somatic mutations derived from 100 representative genes to distinguish cancer types and the results concluded that using somatic point mutations alone as individual variables was not sufficient to classify cancer types. Despite the fact that mutations in many genes have been identified in cancer, it is not yet understood how these genes cumulatively interact in the development and progression of cancer.

Cancer is thought to occur as a consequence of progressive mutational accumulation over time. It has been a challenge to study these mutations and their interactions together due to large-scale complexity. As a result, many groups utilize feature selection as a method for removing irrelevant and redundant information to deal with the complexity problems. Vector Quantization (VQ) [21] and Principle Component Analysis (PCA) [22] have been widely used for feature selection. Recent attention has been drawn to non-negative matrix factorization (NMF). In a face recognition study, Lee et al. presented that NMF outperformed VQ and PCA for feature recognition [23]. In addition, the non-negative constraint of NMF is important because non-negativity is more realistic, easier to interpret, and prevalent in real world applications. NMF has been used for a wide range of applications such as document clustering [24], text mining [24, 25], biological data subgrouping [26], time series analysis [27, 28], blind source separation [29], and signal processing [30]. In particular, NMF has been applied to disease subtype studies using gene expression data [26, 31] and sequencing data [32-34]. However, NMF has not been used as a cancer classifier using somatic mutations obtained from whole-exome sequencing data.

In our study, we proposed a novel method of using NMF to reduce the genetic complexity behind cancer development using somatic mutation information. We hypothesize that the mutations identified individually or in combination will be significantly different between cancers that directly affect processes involved with oncogenesis. Using NMF, we hope to identify latent groups of genes that link to different cancers, aiming at improving the feature selection of genetic mutations by reducing false negatives and at the same time amplify the effects of true positives, thus to improve their statistical power in predicting cancer types.

In this study, NMF and support vector machine (SVM) [35] methods were used to study somatic mutations for multiple cancers. As a pilot study, four cancers were used for subsequent studies, including Glioblastoma Multiforme (GBM), Breast invasive carcinoma (BRCA), Lung Squamous Cell Carcinoma (LUSC), and Prostate Adenocarcinoma (PRAD). NMF was applied to discover latent factors from the somatic mutations within a group of selected genes. The latent factors discovered from NMF were then used to derive features to train an SVM as the classifier for the four different cancer types. The NMF-SVM combination was rigorously evaluated and compared to different baselines. Association studies were performed between the factor matrices derived from NMF and cancer type using penalized logistical regressions. Major factors that are associated with each cancer type were investigated, and significant genes were identified. Using the identified gene sets, enrichment studies were performed using Metascape [36], and the significant GO terms were reported. The details of the study are reported below.

## III. METHODOLOGY

### A. Mutation Profiles

Somatic mutation from four cancers including Glioblastoma Multiforme (GBM), Breast invasive carcinoma (BRCA), Lung Squamous Cell Carcinoma (LUSC), and Prostate Adenocarcinoma (PRAD) were identified from 2431 tumors. The number of samples and mutations varied between cancers as shown in Figure 1. The mutations were functionally annotated by SnpEFF[37] and ANNOVAR [38]. In this study cohort, 245,888 mutations were predicted to have moderate effects by SnpEFF, and 57,319 mutations were predicted to have high effects. Moderate effects are defined as missense mutations and high effects are defined as nonsense mutations. Each mutation was functionally scored using SIFT [25], PolyPhen2 (PP2) [26], and CADD [24] scores as obtained by ANNOVAR. In genes containing multiple mutations, SIFT, PP2, and CADD scores, as well as mutational frequency were collapsed and studied as a single variable separately, which is known as gene burden [39]. These mutation frequency, SIFT, PP2, and CADD scores measure the damage levels of genes from various perspectives. To illustrate the methods used in this study, we have summarized the workflow in Figure 2.

### B. Gene Preselection

Prior to modeling, we preselected a group of representative genes in cancer to achieve a more balanced sample feature ratio and reduce noise. Somatic mutations from 8198 tumors of 18 cancers were retrieved from The Cancer Genome Atlas (TCGA) database. Only mutations in protein-coding genes were used. The number of mutations or collapsed scores in each gene was used as the input variable while the cancer type was used as the output variable. Multinomial logistic regression was fit, and a P-value that yields the null hypothesis of corresponding coefficient being zero was used as an indicator for the pre-selection. The selection criterion for this initial screening was set as P-value less than or equal to a cutoff. In order to reduce noise and to prevent the model from being overfitted, we tested the model using multiple cutoff thresholds of 0.05, 0.1, 0.2, 0.5, and 1. We compared the results derived from each threshold and selected the most reasonable cutoff regarding prediction accuracy and number of features.

| Cancer | | Sample Size | Somatic Moderate | Somatic High |
|---|---|---|---|---|
| GBM | Glioblastoma Multiforme | 763 | 133(55) | 27(9) |
| BRCA | Breast Invasive Carcinoma | 1,044 | 68(11) | 20(5) |
| LUSC | Lung Squamous Cell Carcinoma | 420 | 214(15) | 44(3) |
| PRAD | Prostate Adenocarcinoma | 332 | 34(19) | 8(3) |
| | Total | 2,431 | | |

Figure 1: Sample size in each cancer type. The somatic moderate are the mutations that have moderate effect predicted by SnpEFF. The somatic high are ones predicted to have high effects. The numbers in parentheses are 95% confidence intervals.

C. *Applying NMF to Discover Latent Factors of Somantic Mutations*

Genes passing the selection threshold were used as input for NMF. In the study, we assume that there are N subjects and M selected genes. The data was represented separately by a matrix $A_{Score}$ of size $N \times M$. The columns of $A_{Score}$ represents the collapsed score of the $M$ genes in the $N$ subjects. The matrix $A_{Score}$ was then factorized by NMF. The purpose of this study was to find a set of representative features that are likely to distinguish cancer types.

To perform $NMF$, the matrix $A_{Score}$ was factored into two low-rank matrices $W$ and $H$. Mathematically, $A_{Score}$ is approximated by:

$$A_{Score} \approx W \times H \qquad (1)$$

Matrix $A_{Score}$ is the approximate linear combinations of the column vectors in matrix $W$ and the coefficients supplied by columns in matrix $H$. Matrix $W$ has size $N \times K$, with each of the $K$ columns representing a group of weighted genes and $w_{ij}$ corresponding to the weight of gene $i$ in group $j$. Matrix $H$ has size $K \times N$, where each of the $N$ columns denotes the feature coefficients for each subject. Entry $h_{ij}$ is the value of feature $i$ in sample $j$.

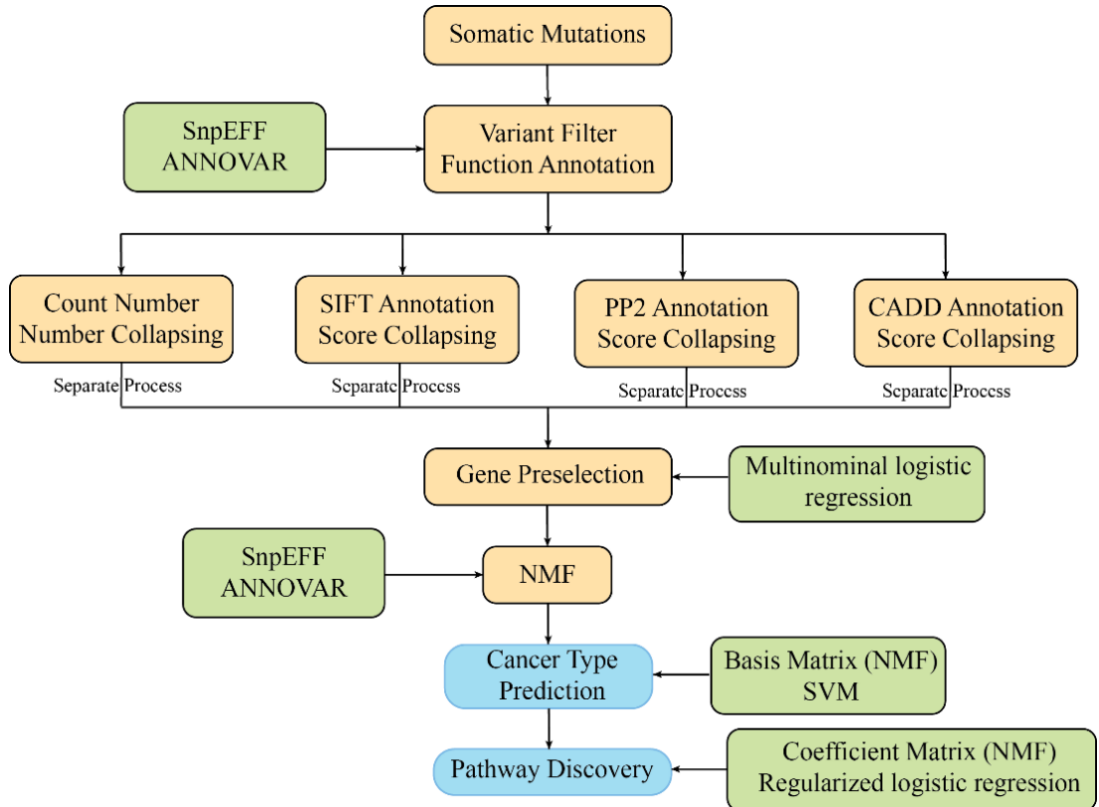

Figure 2: Workflow of the study. Orange boxes are the data or processes; green boxes are the tools used; blue boxes are the results of the study.

The decomposition is achieved by iteratively updating the matrix $W$ and $H$ to minimize a divergence objective [23, 40]. Specifically, for the purpose of sparseness, we used non-smooth Nonnegative Matrix Factorization (nsNMF) for feature selection [41]. Each experiment was repeated ten times to address the local optima problem. Then, the derived features were applied to train a classifier for cancer type.

### D. Classifier Training

Matrix $H$ has size $K \times N$, where each of the $N$ columns denotes the feature coefficients for the corresponding subject. Each sample is placed into the group corresponding to the highest feature coefficient. In this study, to retain the information from both matrices, a new matrix $F$ was generated by multiplying matrix $A_{Score}$ with matrix $W$. Specifically, an entry in matrix $F$ can be computed as:

$$f_{ij} = \sum_{x=1}^{n} A_{ix} * W_{xj} \quad (2)$$

Matrix F has size $N \times K$, $N$ denotes the number of subjects, $K$ denotes the number of factors and is a given input. Each of the $K$ columns represents the accumulative effect of $M$ genes for each subject. Since $A_{Score} \approx W \times H$, matrix $F$ can be approximated by a kernel matrix $W \times H \times W$. In this way, matrix $F$ retains the information from both matrix $W$ and matrix $H$. Subsequently, columns in matrix $F$ can be used as training features to train the classifier. Support vector machine (SVM) was used for the training, each column corresponds to one predictor in the model. The Radial Basis Function (RBF) kernel is used, with parameters of gamma and C set to default. This trained SVM model was then used as a classifier for subsequent cancer type prediction.

### E. Factor Number Selection

Note that before factorization, a pre-defined number of factors $K$ need to be selected. Typically, the number of factors K is chosen so that $(N + M) \times K < N \times M$ [23]. Selection of $K$ is critical because it determines the performance of the classifier. Numerous studies have presented different methods for factor number selection: The factor number $K$ can be determined based on different metrics composing of a cophenetic correlation coefficient [26, 41], variation of sum of squares [42], or maximum information reservation [43]. In our study, the most important feature for the classifier is the ability to distinguish the cancer types correctly. To achieve this goal, a numerical screening test was conducted to screen through the different number of factors for best prediction performance. The screened factor numbers ranged from 2 to 25. Multi-class prediction accuracy in each classification was obtained as a performance measurement. Five-fold cross-validation was conducted using multiple different factor numbers. Each experiment was replicated ten times with different initial seeds. Prediction accuracy, precision, recall, and f-measure were used as performance evaluation matrices.

### F. Evaluation

To set up baselines for comparison, the mutation frequency and the collapsed scores in the selected genes were used as independent predictors to fit a SVM model and a penalized logistic regression model to predict cancer types. All somatic mutations were also used as predictive variables for cancer classification using the SVM model as well as the penalized logistical regression model. These methods were reported to have the best performances for cancer classification using somatic mutations [20]. All the studies were replicated ten times with different initial seeds and significance tests were performed. P-values were obtained for the evaluations.

### G. Pathway Study

In the NMF formulation, the features matrix $F$ was obtained by multiplying matrix $A_{Score}$ with matrix $W$, and it has size $N \times K$, with each $K$ columns representing the accumulative effect of $M$ genes for the corresponding subject. To determine the association level of each factor with each cancer type, we studied each of the K factors' coefficient in associating with the disease type. Elastic net regularization was used for this study. Elastic net regularization penalizes the size of the regression coefficients based on both L1 norm and L2 norm. The estimates from the elastic net methods are defined as:

$$[\widehat{\boldsymbol{\beta}}, \hat{\beta}_0] = argmin_\beta (\sum_i (y_i - \beta' A_{Score} W)^2 + \lambda_1 \sum_{k=1}^{K} |\beta_k| + \lambda_2 \sum_{k=1}^{K} \beta_k^2) \quad (3)$$

$y_i$ denotes the disease type, $\widehat{\boldsymbol{\beta}} = [\hat{\beta}_1, ..., \hat{\beta}_k]$ is the vector of regression coefficients for the K predictors.

In this study, the regulation parameter λ was selected using cross-validation. $\widehat{\boldsymbol{\beta}}$ denotes the association level of each gene group to disease type. W is the weighted coefficient of each gene in the gene groups. We obtained each gene's accumulative effect by multiplying matrix W with vector $\widehat{\boldsymbol{\beta}}$. We used vector $E$ to represent each gene's association effect with each cancer type. Mathematically, let

$$E_i = \sum_{j=1}^{K} \hat{\beta}_j \times W_{ij} \quad (4)$$

denote the $i_{th}$ gene's association effect with the case and control. To note, the matrix $W$ was minmax normalized. Thus, we achieved a score that represents a gene's association effect. To differentially discover pathways for a cancer type, subjects with the cancer type of interest were treated as cases and the rest subjects as controls. After fitting the regression model, we selected the factors corresponding to the largest $\widehat{\boldsymbol{\beta}}$, and denote the factor as a cancer-specific factor. The variables in the factor, are different genes with coefficients. There is a potential that the genes with the largest coefficients are the same genes that cumulatively and linearly interact with each other to cause cancers. For each cancer, we repeated the experiment and selected the top 100 genes in each factor for enrichment analysis and presented the top 2 significant pathways (p < .05).
.

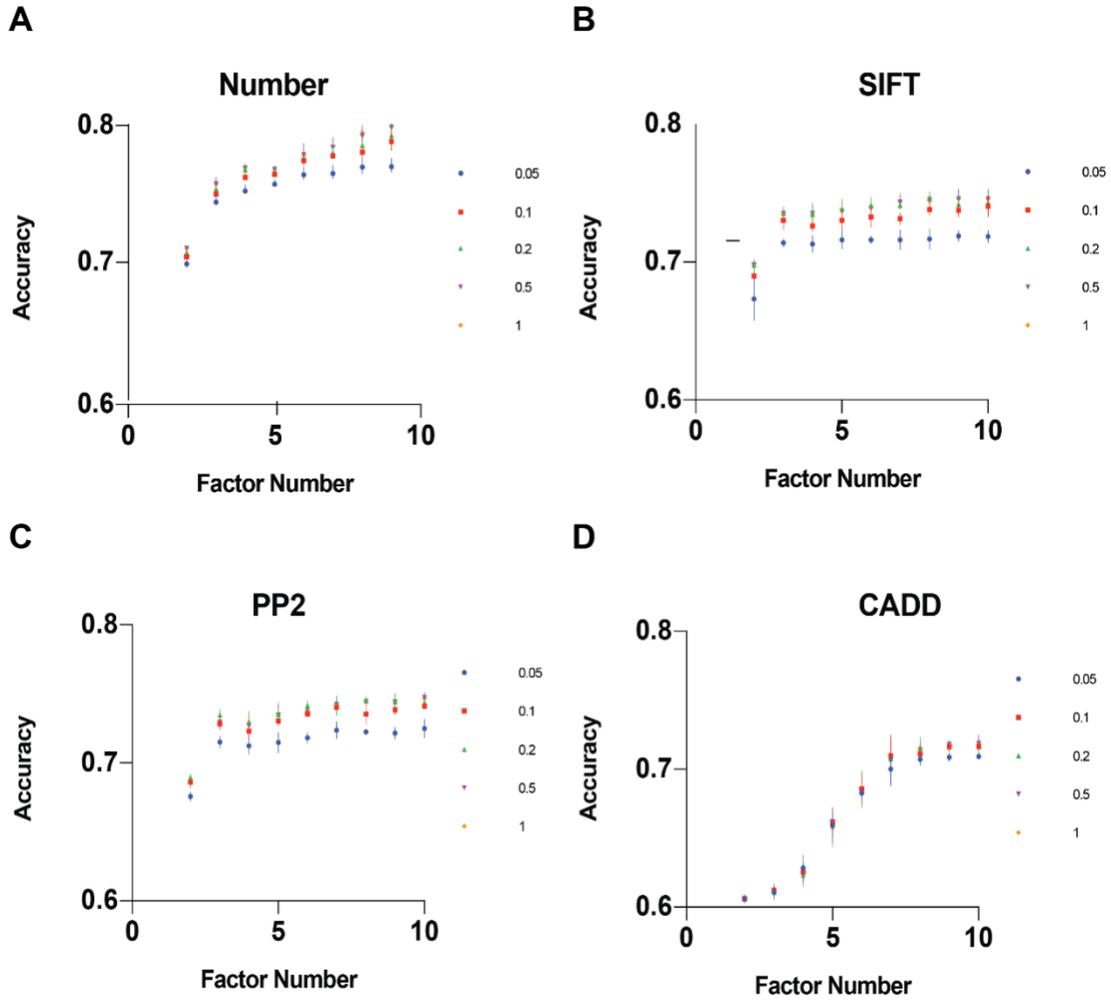

Figure 3: The accuracy of cancer type classification using different P-value cutoffs. (A) Sum of the count of mutations (B) Sum of the SIFT scores (C) Sum of the PP2 scores (D) Sum of the CADD scores.

## IV. EXPERIMENT RESULTS

After collapsing mutations' numbers and scores in each gene, a matrix $A_{score}$ was formed with the 2413 subjects as rows. Entry $A_{ij}$ denotes the $j_{th}$ gene's collapsed score for the $i_{th}$ subject. For the gene preselection, we screened the $p_{value}$ of 0.05, 0.1, 0.2, 0.5, and 1 as cutoffs. For each cutoff, prediction accuracy was used as performance measurements. The experiment was repeated using the Number (sum of the number of mutations), SIFT (sum of the sift scores), PP2 (sum of the PP2 score), and CADD (sum of CADD scores) matrices. Factor numbers ranged from 0 to 10 (Figure 3). Using a cutoff of 1 results in feature numbers 20 times more than using a cutoff of 0.05. However, the performance derived from the cutoff of 1 is not significantly different from the cutoff of 0.05. The $p_{value}$ equal 0.29 comparing the two performances when using Number matrix. Balancing the number of features to be included for computation and accuracy, we selected the cutoff of 0.05 for gene preselections. Following gene preselection, 7,237, 9,483, 8,787, and 8,219 genes were retained for subsequent analysis for the Number, SIFT, PP2, and CADD matrix respectively.

We then compared these four matrices: Number, SIFT, PP2, and CADD. The number of factors K ranged from 2 to 25. This range is within the constraint of the rule (N+M)K<NM. For each factor number in this range, nsNMF was applied to the matrices, and the corresponding classifier was trained. The performances derived from the matrix Number outperformed the other matrices significantly (Figure 4). The precision, recall, and f-measures were also derived and similar patterns and trends were observed. Given the better performance, the matrix Number was used for subsequent analyses. Using Number matrix, the maximum accuracy is 78.4% (Standard Error of the Mean SEM=0.2%) when the factor number equaled 21. For performance accuracy, there is an inflection point when the cluster number increase from 10 to 11 (Figure 4). The accuracy becomes stable when factor number is larger than 10. To prevent the potential overfitting, we chose the factor of 10 for our analysis. With this factor number, the overall accuracy is 77.1% (SEM=0.1%).

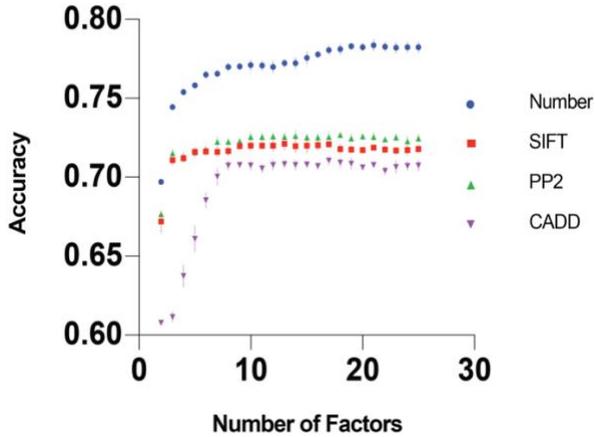

Figure 4: The accuracy of cancer type predictions using different numbers of factor. The four plots were the accuracies derived from the matrices of Number, SIFT, PP2, and CADD scores.

The performance of our proposed model (77.1%, SEM=0.1%) significantly outperformed the other four baselines (Figure 5). The P-value for Student's t-test was 0.002 comparing our proposed model to the second-ranked model (73.9% (SEM=0.8%), which applies penalized logistical regression with the aggregated Number matrix. In the baselines, aggregating the mutations in a gene has improved the performance significantly too (p<0.01 in both comparisons).

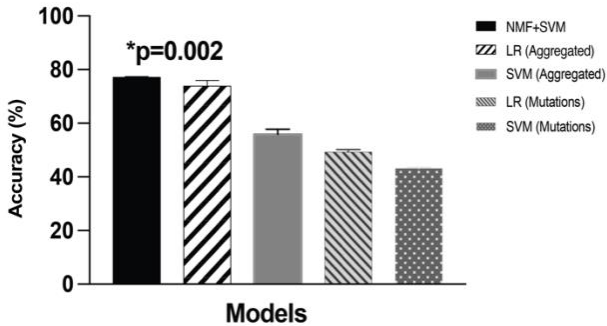

Figure 5 Comparison of our proposed model (nsNMF+SVM) with baselines. LR is penalized logistical regression. SVM is support vector machine. Aggregated is the matrix to sum variants together in the same gene. Mutation is the model that utilizes every single mutation as an input variable.

In the regularized logistical regression study, we assessed each gene's association effect with each cancer type. From the formula (4), the association score is the sum of the feature weight multiplied by the weights of each gene in each feature. High scores indicate a significant role of a mutated gene in disease. We selected ten genes with the top association scores for breast cancer and found many of them have been previously implicated in breast cancer. *MUC16* and *RYR3* have been shown to play a role in tumorigenesis and have enriched expression in breast cancer patients [44-46]. Mutations in *USH2A*, and *CTNNA* have been identified to be associated with breast cancer risk in independent studies [47, 48] The nebulin protein family, *HUWE1*, and *COL22A1* are also important genes that have been implicated in other cancers [49-51].These genes are summarized in Table 1.

TABLE 1: THE GENE LIST WITH HIGHEST COEFFICIENTS FOR BREAST CANCER

| Gene Symbol | Score | Description |
|---|---|---|
| CSMD3 | 0.63 | CUB and Sushi multiple domains 3 |
| NEB | 0.59 | nebulin protein coding gene |
| MUC16 | 0.56 | mucin 16 |
| HUWE1 | 0.50 | HECT, UBA and WWE domain containing 1 |
| USH2A | 0.41 | usherin protein coding gene |
| CTNNA2 | 0.40 | catenin (cadherin associated protein), alpha 2 |
| DNAH6 | 0.39 | dynein, axonemal, heavy chain 6 |
| MYHAS | 0.37 | myosin heavy chain gene antisense RNA |
| RYR3 | 0.36 | ryanodine receptor 3 |
| COL22A1 | 0.35 | Collagen type XXII alpha 1 chain |

The top 100 genes associated with each cancer type were derived and analyzed in Metascape for pathway enrichment. Overall, we identified pathways that have been directly linked to cancer such as morphogenesis and ion transport pathways [52-54]. Furthermore, we also identified and confirmed some pathways known to be associated with specific cancers such as AKT signaling and glioblastoma (GBM) cancer [55, 56]. The top 2 significant pathways ($p < .05$) for each cancer type are illustrated in TABLE 2.

TABLE 2: BIOLOGICAL PROCESSES MOST ASSOCIATED WITH EACH CANCER TYPE.

| GO Term | Cancer |
|---|---|
| Microtubule-based process | BRCA |
| Calcium ion transmembrane transport | BRCA |
| Focal adhesion | GBM |
| PI3K-Akt signaling pathway | GBM |
| Maintenance of protein location | PRAD |
| Cytoskeletal anchoring at plasma membrane | PRAD |
| Neuron projection morphogenesis | LUSC |
| Plasma membrane bounded cell projection morphogenesis | LUSC |

## V. DISCUSSION

In this study, we have proposed a method to use somatic mutations to classify the cancer type and to derive the relevant genes and pathways. This is a new method to understand the somatic mutations. We used 8,198 tumors' somatic mutational information to select 7,237 cancer-relevant genes. We then applied nsNMF and SVM to train a classifier to distinguish the 2,413 tumors among four cancer types including Glioblastoma Multiforme (GBM), Breast invasive carcinoma (BRCA), Lung Squamous Cell Carcinoma (LUSC), and Prostate Adenocarcinoma (PRAD). Products of basis matrix and coefficient matrix derived from nsNMF were both retained to construct the feature matrix. Subsequently, the constructed features were used as input variables to train the classifier. We compared functional scores using CADD, SIFT, and PP2, and counted mutation number and found that counted mutation number yielded the best performance (accuracy=77.1% with SEM=0.1%). Finally, regularized logistical regression was applied to study each gene's association effect with cancer type. Using the associated features, we derived relevant genes and pathways for each cancer.

When training the classifier, we used an alternative method by multiplying the matrix $A_{score}$ with matrix $W$ to obtain the feature matrix $F$. Thus, information from basis component W was retained, providing information about weights in each gene group. This information was then used as features to train the classifier. Another benefit of this alternative method is the resultant ease at the testing stage. With the trained $W$ matrix, we only need to left multiply the testing $A_{score}$ matrix in order to get the test feature matrix. In addition to improving cancer type classification, each gene's association effect with the cancers was of interest and also studied.

The development of high throughput sequencing technology has enabled the cataloging of large-scale mutation information. Traditionally, mutations derived from sequence data were examined as a single variable using the regression models [20, 57]. Unfortunately, the large number of variables have limited the power of such studies. To reduce the number of variables, studies have proposed to aggregate mutations at the gene level as input for the regression model [39, 58, 59]. In other studies, mutations in a gene have also been proposed to be studied in a matrix as input for kernel test [60, 61]. In this study, we have proposed a new framework, which utilizes a regression model to preselect deleterious genes, nsNMF to decompose the matrix, SVM to train a classifier, and then penalized regression to derive relevant genes. In this framework, we have carefully tuned the parameters and models. We have also proved that this is an effective model to classify cancers, to derive relevant genes, and to study pathways.

Given that the preselection process includes all our analyzed four cancers' datasets, there is a potential of overfitting. We have used available data from all 18 cancers to reduce the potential overfitting issues. Nevertheless, our study still faces some limitations from overfitting and additional datasets will need to be included.

## VI. CONCLUSIONS

To fully understand a disease, studying a full range of genes together is of critical importance. In addition, complex traits are modified by multiple genes and multiple mutations together [62]. Traditionally, NMF has been applied to study gene expression[26, 43]. In this study, we proposed using the somatic mutations for cancer classification. Furthermore, we proposed generating the feature matrix by integrating both the basis matrix W and the coefficient matrix H. Moreover, we developed a method to derive effect scores from the feature matrix. Using this method, we obtained the association score of each gene with a particular cancer type, which then enabled us to discover potential relevant pathways. The discovered effect scores have a high potential to help us better understand genetic pathophysiology behind cancer.

In this study, we propose a novel strategy to study the difference of the genetic landscape of cancer. In the future, we will use tensor factorization to integrate known pathways to guide the grouping of mutational variants [63] and use external cohorts to validate the proposed model. Furthermore, this generic process only requires the input of somatic mutations and disease type of interest, without much domain specific knowledge. This strategy has the potential to be easily adapted and applied to other diseases as well.